\documentclass[twocolumn,showpacs,showkeys,preprintnumbers,amsmath,amssymb]{revtex4}
\usepackage{graphicx}% Include figure files
\usepackage{dcolumn}% Align table columns on decimal point
\usepackage{bm}% bold math

\begin{document}

\title{Full $fp$-shell Study of Even-Even $^{48-56}$Ti Isotopes}

\author{F.~A.~Majeed$^{1,2}$} \email{fal-ajee@ictp.it}
\author{A.~A.~Auda$^3$}
\affiliation{$^1$Department of Physics, College of Science,
Al-Nahrain University, Baghdad, Iraq} \affiliation{$^2$The Abdul
Salam International Centre for Theoretical Physics}
\affiliation{$^3$Department of Physics, College of Teachers,
Al-Jabal Al-Gharby University, Gharyan, Libya}

\date{\today}

\begin{abstract}
The level schemes and transition rates {\em B}({\em
E}2;$\uparrow$) of eve-even $^{48-56}$Ti isotopes were studied by
performing large-scale shell model calculations with FPD6 and
GXPF1 effective interactions. Excellent agreement were obtained by
comparing the first 2$^{+}$ level for all isotopes with the
recently available experimental data, but studying the transition
strengths {\em B}({\em E}2; 0$^+_{g.s.} \rightarrow$2$^+_1$) for
all Ti isotopes using constant proton-neutron effective charges
prove the limitations of the present large-scale calculations to
reproduce the experiment in detail.
\end{abstract}

\maketitle

\section{Introduction}
The structure of neutron-rich nuclei has recently become the focus
of much theoretical and experimental effort. Central to the
on-going investigation is the expectation that substantial
modifications can occur to the intrinsic shell structure of nuclei
with a sizable neutron excess \cite{BA01}.

 Interactions between protons and neutrons have been also invoked to account for the
presence of a sub-shell gap at N=32 in neutron-rich nuclei located
in the vicinity of the doubly-magic nucleus $^{48}$Ca \cite{JI01}.

Full $pf$-shell model study of A=48 nuclei were performed by
Caurier and Zuker \cite{EC94} by modifying Kuo-Brown (KB)
\cite{KB68} to KB1 and KB3. The isobaric chains A=50, A=51 and
A=52 studied by Poves {\em et al.} \cite{AP01} using KB3 and FPD6
\cite{WR91} and their new released version KB3G.

Reduced transition probabilities to the first 2$^{+}$ state in
$^{52,54,56}$Ti and the development of shell closure at N=32, 34
were studied by Dinca {\em et al.} \cite{DD05} both experimentally
and theoretically using the most recently modified interaction
labeled GXPF1A done by Honma {\em et al.} \cite{HM05}. They
confirm the presence of a sub-shell closure at neutron number N=32
in neutron-rich Ti nuclei above $^{48}$Ca and this observation are
in agreement with the shell model calculations using the most
recent effective interaction, also they conclude that the data do
not provide any direct indication of the presence of additional
N=34 sub-shell gap in the Ti isotopes and that the measured {\em
B}({\em E}2; 0$^+_{g.s.} \rightarrow$2$^+_1$) probabilities
highlight the limitations of the present large-scale calculations
as they are unable to reproduce in detail the magnitude of the
transition rates in semi-magic nuclei and their strong variation
across the neutron-rich Ti isotopes.

The purpose of this letter is to study the reduced transition
probabilities and level schemes of even-even $^{48-56}$Ti isotopes
using the new version of OXBASH for windows \cite{OX05}. The level
schemes of selected states of $^{54}$Ti and $^{56}$Ti calculated
in this work compared with the most recently available
experimental data and with the previous theoretical work in
Ref.\cite{HM05} using GXPF1A, GXPF1 and KB3G interactions.

\section{Shell Model Calculations}
The calculations were carried out in the D3F7 model space with the
FPD6 Hamiltonian \cite{WR91} using the code OXBASH \cite{OX05} for
$^{48}$Ti, while F7P3 model space employed with effective
interaction FPD6 for $^{50}$Ti.

For $^{48}$Ti the core is considered as $^{32}$S with 16 nucleons
outside core, while for $^{50}$Ti the core was taken as $^{40}$Ca
and 10 nucleons outside the core.

The core was taken as $^{48}$Ca for the three nuclei $^{52}$Ti,
$^{54}$Ti and $^{56}$Ti and the model space is (HO) with FPD6
effective interaction. The effective interaction GXPF1 \cite{GX02}
were used also to calculate the level spectra for $^{54}$Ti and
$^{56}$Ti for the purpose of comparison with Ref.\cite{HM05}.
\section{Results and Discussion}
The test of success of large-scale shell model calculations is the
predication of the first 2$^{+}$ level and the transition rates
{\em B}({\em E}2; 0$^+_{g.s.} \rightarrow$2$^+_1$) using the
optimized effective interactions for the description of {\em
fp}-shell nuclei.

Figure 1 presents the comparison of the calculated {\em
E}$_x$(2${^+_1}$) energies with FPD6 from the present  work with
the experiment, the work of Dinca {\em et al.}\cite{DD05} and with
the most recent calculations using the new effective interaction
labeled GXPF1A \cite{DC05}. The comparison shows that FPD6
effective interaction is better than GXPF1 except for $^{54}$Ti at
N=32 shell closure, GXPF1 is better in reproducing the {\em
E}$_x$(2${^+_1}$) level. The modified effective interaction GXPF1A
is more successful in description of all the mass region A=48-56,
but only at N=32 shell gap GXPF1 is more successful in reproducing
{\em E}$_x$(2${^+_1}$) for $^{54}$Ti.

The new effective interaction GXPF1A which is the improved type of
GXPF1 are the most convenient one for the whole chain of Ti
isotopes for the mass region A=48-56, but still can not reproduce
the shell gap at N=32 like GXPF1. Our work is also fail to
reproduce the shell gap at N=32.
\begin{figure}
\centering
\includegraphics[width=0.44\textwidth]{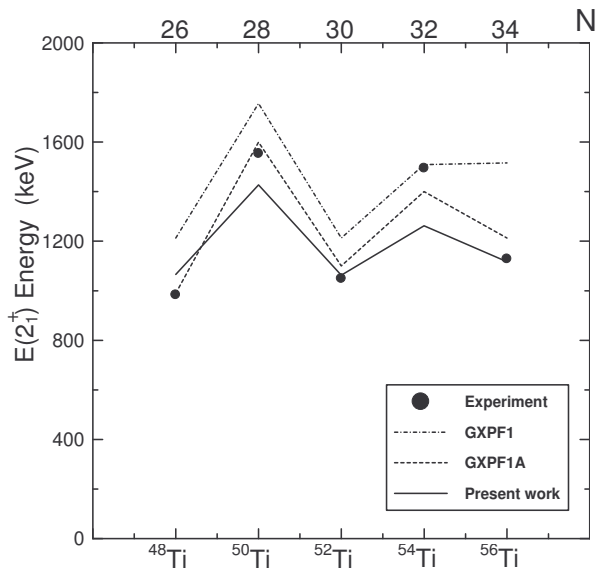}
\caption{Systematics of {\em E}$_x$(2${^+_1}$) for eve-even Ti
isotopes. Experimental data (closed circles) are compared with
present work (solid line), the previous work using GXPF1
(dashed-dot-line) and GXPF1A (dashed line). Experimental data are
taken from Refs.\cite{BF04,RE00}.}
\end{figure}

Figure 2 shows the large-scale shell model calculations of the
reduced transition strengths {\em B}({\em E}2; 0$^+_{g.s.}
\rightarrow$2$^+_1$) that have been performed by adopting the
effective charges for proton is e$_p$=1.15$e$  and for neutron
e$_n$=0.8$e$ as suggested in Ref.\cite{SN04} and also these values
were used in the calculations of the previous work using GXPF1 and
GXPF1A in Ref.\cite{DC05}.

The solid line in Fig.2 is the present calculations using the
effective interaction FPD6 compared with the most recently
measured experimental data and with the previous work using GXPF1
and the new modified interaction GXPF1A. Our calculations produced
staggering in the calculation of B(E2) and it is in better
agreement with experimental data as compared with the previous
theoretical work \cite{DD05} even when they choose the modified
interaction GXPF1A, but our work compared with the recent
theoretical work of Poves {\em et al.} \cite{AF05} their
calculations using KB3G effective interaction are in better
agreement with the experiment for the nuclei $^{48, 50, 54,
56}$Ti, but not $^{52}$Ti at N=30 our results are in better
agreement with experiment. Although that GXPF1A effective
interaction is in better agreement in reproducing the first
2$^{+}$ level in all even-even Ti isotopes for the mass region
A=48-56 but still not able in reproducing the experimental data
for the {\em B}({\em E}2; 0$^+_{g.s.} \rightarrow$2$^+_1$)
transition strengths. The difference between our calculations and
the previous theoretical work from Ref.\cite{DC05} is mainly
attributed to the difference of the location of the
single-particle energies $f$$_{7/2}$, $p$$_{3/2}$ , $f$$_{5/2}$
and $p$$_{1/2}$ for the effective interactions FPD6, GXPF1 and the
modified one GXPF1A which effect significantly the predication of
level excitations and transition strengths B(E2).
\begin{figure}
\centering
\includegraphics[width=0.44\textwidth]{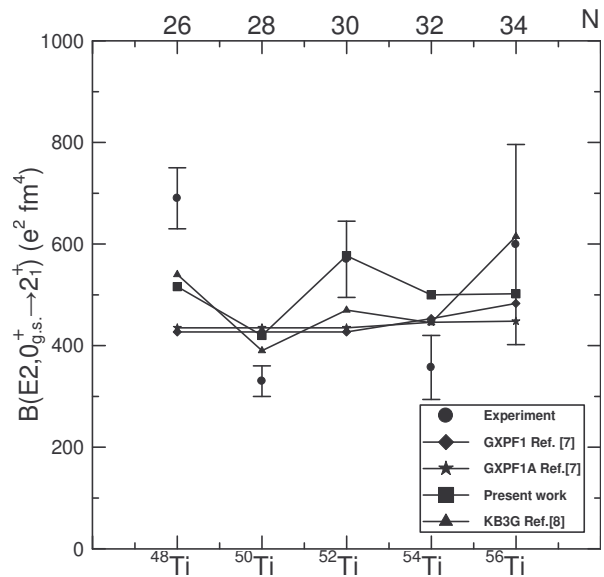}
\caption{Comparison of the large-scale shell model calculations
using FPD6 (squares)   with the experimental {\em B}({\em E}2;
0$^+_{g.s.} \rightarrow$2$^+_1$) transition strengths (closed
circles) for the chain of even-even Ti isotopes and with the
previous work using the effective interactions GXPF1 (diamonds)
and GXPF1A (stars) and with the work from Ref.\cite{AF05} using
KB3G effective interaction. Experimental data are taken from
Refs.\cite{DC05,JW81}.}
\end{figure}

The calculated FPD6 and GXPF1 energy levels are compared with the
experimental data and the previous work using GXPF1A, GXPF1 and
KB3G as shown in Fig. 3. The agreement is excellent for
$J$$^{\pi}$ =0$^{+}$, 2$^{+}$, 4$^{+}$ and 6$^{+}$ sequence with
FPD6 effective interaction. In order to improve the description of
{\em E}$_x$(2${^+_1}$) for $^{56}$Ti, one possible choice is to
lower the single particle energy of the $f$$_{5/2}$ orbit by 0.8
MeV, as suggested  in Ref.\cite{SN04}.

The reduction of $f$$_{5/2}$ orbit by 0.8 MeV improve the
prediction of {\em E}$_x$(2${^+_1}$) as shown in Fig. 4 for
$^{56}$Ti and it remedies this discrepancy by about 0.2 MeV.
However, such a modification improve the prediction of {\em
E}$_x$(2${^+_1}$) in $^{54}$Ti also, but it is fail completely in
description of high spin states.
\begin{figure}
\centering
\includegraphics[width=0.38\textwidth]{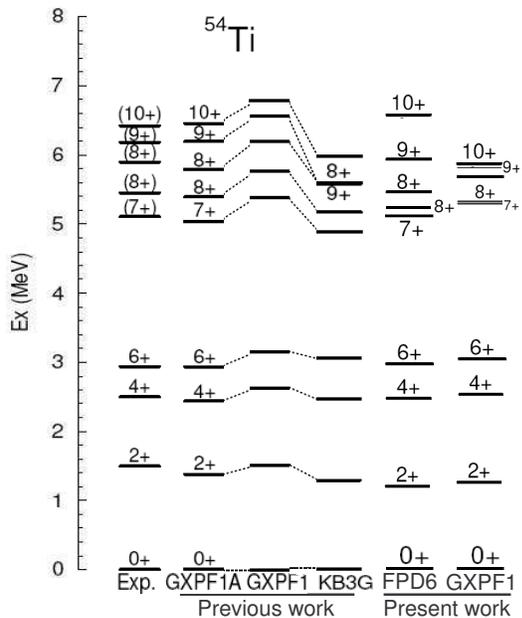}
\caption{Comparisons between shell-model calculations with FPD6
and GXPF1 effective interactions (present work) with the
experimental energy levels for the positive parity states of
$^{54}$Ti and with the theoretical work using GXPF1, KB3G and
GXPF1A (previous work) Ref.\cite{HM05}. Experimental data are
taken from Ref.\cite{RV02}.}
\end{figure}

It can be seen in Fig.4 that GXPF1 predicts {\em E}$_x$(2${^+_1}$)
better than FPD6 and almost its prediction as compared with
previous work using GXPF1A is excellent, but it is not good in
description of high spin states of $^{56}$Ti and still FPD6 is in
better agreement in describing the high spin states. Besides FPD6
predicts the level sequence $J$$^{\pi}$=8$^{+}$, 7$^{+}$ ,
9$^{+}$, while GXPF1 predicts $J$$^{\pi}$=9$^{+}$, 8$^{+}$,
7$^{+}$.
\begin{figure}
\centering
\includegraphics[width=0.36\textwidth]{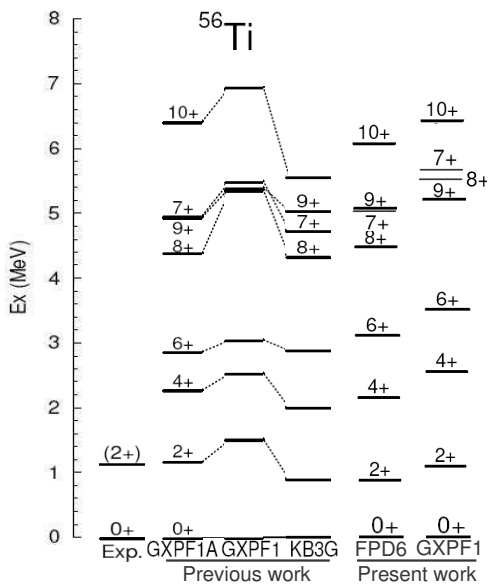}
\caption{Calculated energy levels of $^{56}$Ti with two effective
interactions FPD6 and GXPF1 compared with the experimental data
and with the previous theoretical work using GXPF1, KB3G and
GXPF1A Ref.\cite{HM05}. Experimental data are taken from
Ref.\cite{SN04}.}
\end{figure}
\newpage
\section{Summary}
Large-scale shell model calculations by adopting FPD6 and GXPF1
effective interactions were used to calculate the level excitation
and transition strengths {\em B}({\em E}2; 0$^+_{g.s.}
\rightarrow$2$^+_1$) for the mass region A=48-56 for the even-even
Ti isotopes. The comparison of the calculated {\em B}({\em
E}2;0$^+_{g.s.} \rightarrow$2$^+_1$) with the measured
experimental data even with the small staggering prove the
conclusions made by Refs.\cite{DD05,AF05} that there is
limitations of the present large-scale calculations to reproduce
in detail the magnitude of the transition rates in the semi-magic
nuclei and their strong variation across the neutron-rich Ti
isotopes.
\section{Acknowledgments}
The first author F. A. M. would like to acknowledge the The Abdus
Salam International Centre for Theoretical Physics (ICTP) for the
financial support and warm hospitality.
\newpage

%\bibliography{apssamp}

\end{document}